\documentclass[]{article}
\usepackage{lscape}
\usepackage{pdflscape}
\usepackage{graphicx}
\usepackage{multirow}
\usepackage{dcolumn}
\usepackage{natbib}
\usepackage{amsmath}
\usepackage{amssymb}
\usepackage{listings}
\usepackage[T1]{fontenc}
\usepackage[utf8]{inputenc}
\usepackage{graphicx}
\usepackage{epstopdf}
\usepackage{color}
\usepackage{float}
\usepackage{tikz}
\usepackage{tcolorbox}
\usepackage{multirow}
\usepackage{hhline}
\usepackage{xcolor}
\usepackage{colortbl}
\usepackage{subcaption}
\usepackage{booktabs}
\usepackage{lscape}
\usepackage{subcaption}
\usepackage{pgfplots}
\usepackage{lineno}
\usepackage[utf8]{inputenc}
\usepackage[letterpaper,pdftex]{geometry}
\usepackage{longtable}
\usetikzlibrary{arrows,automata,fit,scopes,calc,matrix,positioning,decorations.pathmorphing,decorations.pathreplacing}
\pgfplotsset{width=9cm,compat=1.15}
\usepgfplotslibrary{statistics}
\usepackage[all]{xy}
\usepackage[affil-it]{authblk}
\usepackage[doublespacing]{setspace}
\usepackage{verbatim}
\usepackage{rotating}

\usepackage{hyperref}	
\hypersetup{colorlinks,linkcolor={blue},citecolor={blue!50!black},urlcolor={blue}}
\usepackage{geometry}

\title{Home advantage and crowd attendance: Evidence from rugby during the Covid 19 pandemic}
\author[1,2]{Fernando Delbianco \footnote{fernando.delbianco@uns.edu.ar}} 
\author[1,3]{Federico Fioravanti \footnote{federico.fioravanti9@gmail.com}}
\author[1,2]{Fernando Tohm\'e \footnote{ftohme@criba.edu.ar}}

\affil[1]{Instituto de Matemática de Bahía Blanca, CONICET - UNS, Bah\'{\i}a Blanca, Argentina}
\affil[2]{Departamento de Economía, Universidad Nacional del Sur, Bahía Blanca, Argentina}
\affil[3]{Departamento de Matemática, Universidad Nacional del Sur, Bahía Blanca, Argentina}
\date{}
\begin{document}

\maketitle

\begin{abstract}
 The COVID-19 pandemic forced almost all professional and amateur sports to be played without attending crowds. Thus, it induced a large-scale natural experiment on the impact of social pressure on decision making and behavior in sports fields. Using a data set of 1027 rugby union matches from 11 tournaments in 10 countries, we find that home teams have won less matches and their point difference decreased during the pandemics, shedding light on the impact of crowd attendance on the {\em home advantage} of sports teams.
 
\end{abstract}

\section{Introduction}
A traditional song chanted by fans in many sport fields in support of their teams says \textquotedblleft You'll never walk alone\textquotedblright\footnote{A song composed by Richard Rodgers and sung by Oscar Hammerstein II.}. The crowds chant it to encourage their teams, believing that this will give them the strength to beat the opponent. The idea is that a strong support by the fan base at the stadium can lead to a victory. It is natural to think that such support will be stronger if the game is played at the home stadium. In fact, several studies confirm that home teams have an advantage (known in the literature as {\em home advantage} (HA)) over the away team. It has been furthermore shown that HA leads to a higher winning rate or at least to a larger point difference. While the encouragement of the fan base may be part of it, a full explanation of this phenomenon requires to take into account other factors, like the knowledge that the home team has of the field, the fatigue that the away team may feel for having traveled to get to the home team's stadium, or even the social pressure that the local fans exert over the referees on the field. Schwartz and Barsky (1977), in what is considered the first empirical investigation of the extent of and reasons for HA, suggested that crowds exert an invigorating, motivational influence, encouraging the home side to perform well. After their work, many other researchers investigated HA, with different approaches such as physiological (Neave and Wolfson 2003), psychological (Agnew and Carron 1994, Legaz-Arrese et al. 2013), economical (Carmichael and Thomas 2005, Boudreaux et al. 2015, Ponzo and Scoppa 2018) and even exploring possible referee biases favoring home teams (Downward and Jones 2007, Page and Page 2010). These studies have been replicated for several sports. A vast literature can be found about the home advantage in soccer (Boyko et al. 2007, Taylor et al. 2008, Belchior 2020). HA has been analyzed for other sports as well: basketball (Pojskić et al. 2011, Jones 2007), tennis (Nevill et al. 1997), athletics (Cutcheon 1984, Jamieson 2010), handball (Aguilar et al. 2014), judo (Ferreira et al. 2013, Krumer 2017). This phenomenon has been studied even in the case of major events such as the Olympic games and the soccer World Cup (Balmer et al. 2001 and 2003, Brown Jr et al. 2002).\\

 There exists also a vast literature on HA in rugby union and rugby league. García et al. (2013) found that the HA for tournaments where the top European representatives\footnote{England, Ireland, Wales, Scotland, France and Italy} play amounts to a $60\%$ of win rate for home teams. Kerr and van Schaik (1995) showed that game venue affected the post-game mood (related to arousal) of Dutch rugby male players. Jones et al. (2007) found that in rugby league competitions the cost of behaving aggressively (in terms of game outcome) may be greater for the away team. Page and Page (2010) found that in the Super Rugby tournament\footnote{A tournament played by the best teams from the Southern hemisphere. Due to the Covid restrictions, this tournament will be played only by the New Zealand and Australia teams during the 2021 season.} the home team wins $71\%$ of its matches when the referee has the same nationality and $50\%$ otherwise.  \\

The Covid 19 pandemic created the conditions for a natural experiment to test the impact of local supporters on home advantage. In mid-March of 2020 almost all sports competitions were suspended worldwide, re-starting the season later in the year at different months depending on the country. But this reboot was allowed only on the condition of accepting several new restrictions, such as not letting in large supporting crowds, reducing the attendance capacity or even imposing hygienic protocols that discouraged people from attending in the field. We investigate the impact of these restrictions on different rugby club tournaments  as well as competitions among  national representative teams. Our research contributes to the new increasing literature of HA during the Covid 19 pandemic. To our knowledge, this is the first work that focus on HA in rugby during this situation, as most of them are in soccer (Cueva 2020, Bryson et al. 2021, Endrich and Gesche 2020, McCarrick et al. 2020).\\

In this paper we examine variables representing HA and see how they have been affected by the different restrictions imposed on the attendance to matches in various rugby tournaments. In section 2 we discuss the database used in our study as well as the methodology to assess the possible impact of the restrictions on HA. Section 3 presents the results and section 4 concludes.\\

\section{Data and Methodology}
Our main data set consists of 1027 matches from 11 professional and semi-professional tournaments where 462 games were played before the suspension of all sport activities and 565 after it. We include games from season 2019/2020/2021, both at the end and beginning of the seasons, as well as full tournaments. During these seasons, no major changes were introduced in the rugby law book that could have affected directly the outcomes\footnote{A new tool for a captain's team has been introduced in the 2021 Super Rugby Aotearoa season and other tournaments, called Captain's Challenge, that allows to ask the referee to review a foul play or try situation \url{https://www.rugbypass.com/news/captains-challenge-among-the-latest-law-innovations-adopted-by-super-rugby/}. No game in our database corresponds to any of those tournaments. }. Almost all games played after the suspension respected some crowd restrictions or did not let any public in at all. One exception is, for example, the 2020 Super Rugby Aotearoa in New Zealand that started the competition in June with free access to the public\footnote{Only one game was suspended and one game was played without attendance during the final round due to new Covid restrictions. \url{https://super.rugby/superrugby/news/blues-v-crusaders-match-cancelled/}}.\\
Each observation in this data set consists of a specific match between two rugby teams. Each entry consists of the following items of information: the final score (\textit{PointsL} and \textit{PointsV}); whether the home team did not win (\textit{NotHomeW}); the points difference in the score (\textit{DifLoc}); whether it was played before or during the Covid pandemics (\textit{Covid}); if the restriction consisted in not allowing attendance, allowing a restricted one or permitting free attendance (\textit{Attendance}); whether any kind of restriction was applied at all (\textit{Restricted}); if the full lockdown was enforced at the time (\textit{Closed}); if a win was predicted for the home team \footnote{According to predictions in \url{https://rugby4cast.com/}.}(\textit{Predic}); the final positions of the home and away teams in the previous tournament (\textit{PosL} and \textit{PosV}); the streak of home and away wins (\textit{StreakL} and \textit{StreakV}); whether it was a male or female tournament (\textit{Gender}); the amount of wins per team in the current competition (\textit{WinL} and \textit{WinV}); to which tournament the game corresponded (\textit{Tournament}); the country where the game was played (\textit{Country}) and if the away team had to travel abroad to play the game (\textit{Travel}). The tournaments considered are: Top 14 and Pro D2 (France), Mitre 10 Cup (New Zealand), Premiership and Premier's 15 (England), Currie Cup (South Africa), Super Rugby (Southern Hemisphere), Six Nations, Pro 14 and Champions Cup (Europe).\footnote{Descriptive Statistics can be found in Appendix 1.}\\

\begin{figure}[hbt!]
	\begin{subfigure}{.8\textwidth}
	\centering
	\caption{Closed=0}
	\includegraphics[width=\linewidth]{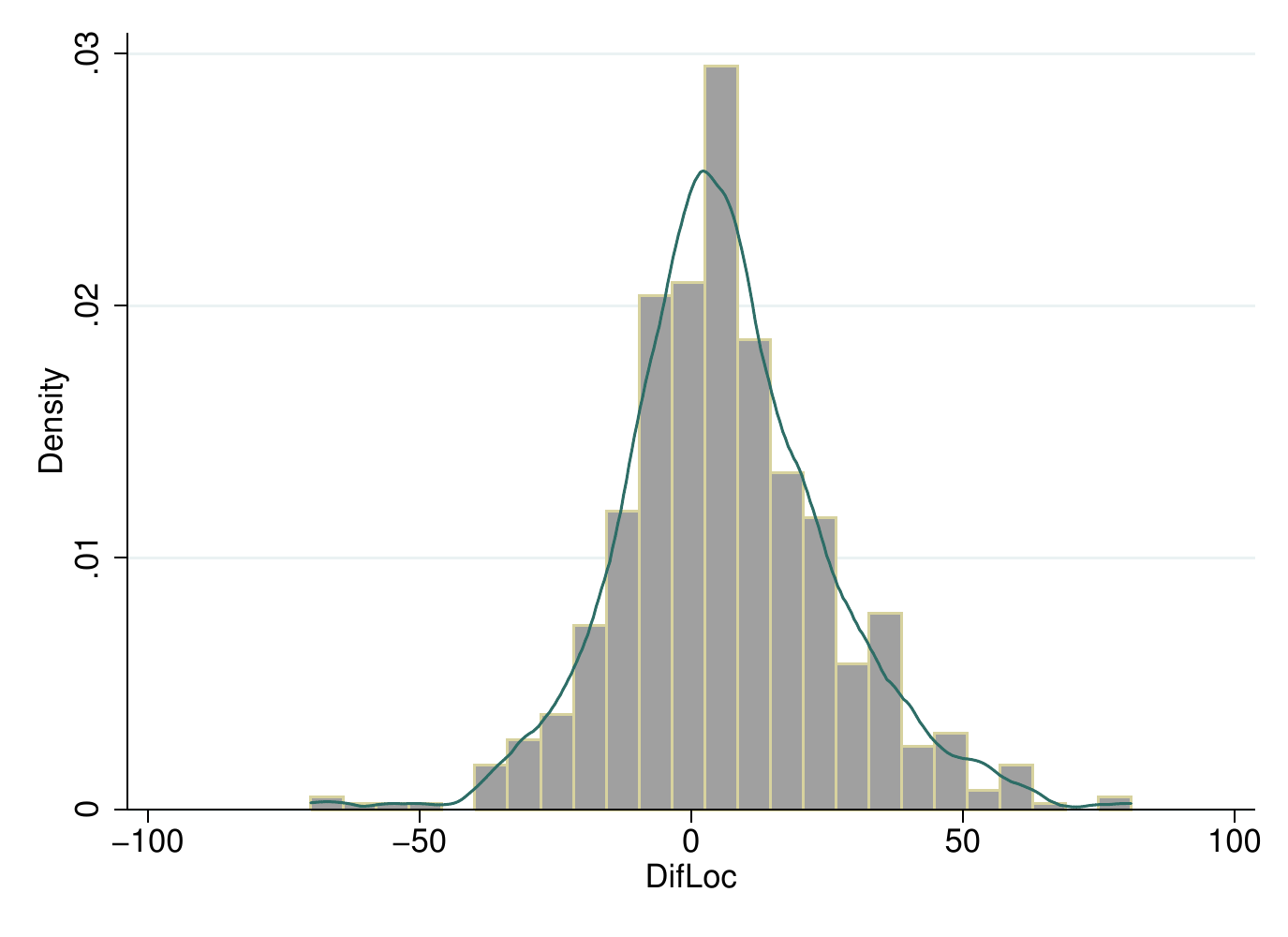}
	\label{fig:cerr0}
\end{subfigure}
\newline
\begin{subfigure}{.8\textwidth}
\centering
\caption{Closed=1}
\includegraphics[width=\linewidth]{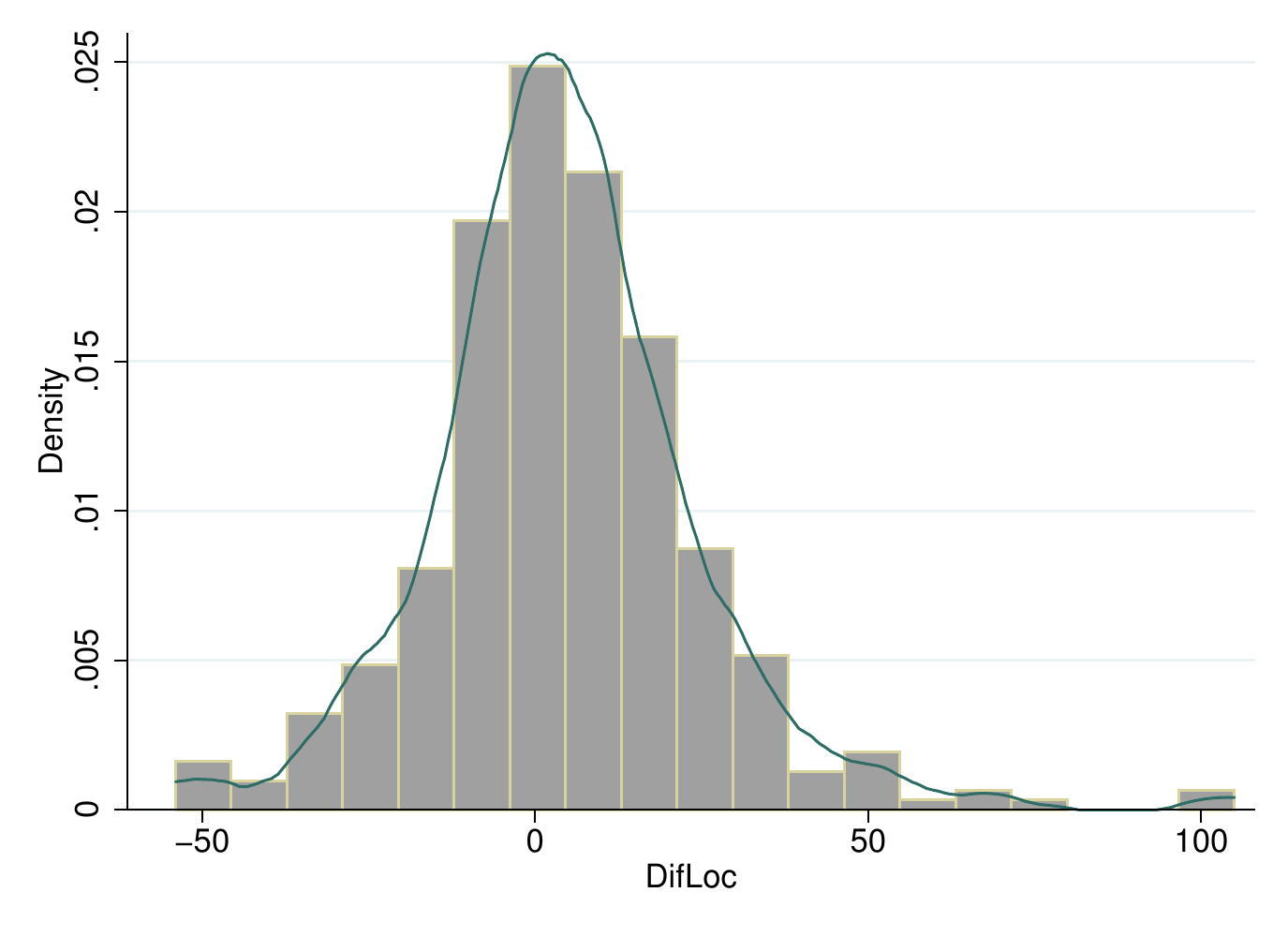}
\label{fig:cerr1}
\end{subfigure}	
\end{figure}

We can see in the histograms \ref{fig:covid0}, \ref{fig:covid1}, \ref{fig:cerr0} and \ref{fig:cerr1} the difference between points scored by the home and away team. We separate the figures depending on whether the data corresponds to games played before or during the Covid pandemics (\ref{fig:covid0} and \ref{fig:covid1}), or depending on if restrictions were imposed or not (\ref{fig:cerr0} and \ref{fig:cerr1})\footnote{These histograms can be found in Appendix 2.}. In the cases before Covid or in absence of  restrictions, we can see an \textquotedblleft outlier peak\textquotedblright{} for narrow home victories, where the host team won by a difference between one and three points. When we consider the Covid or the restricted case (\textit{Covid=1} and \textit{Closed=1}), this \textquotedblleft peak\textquotedblright{} disappears. \\

Given the evidence obtained from the descriptive statistics, we can run two tests. First, a difference of means test to check whether the differences found in the histograms result in significantly different first moments of the distribution. Then, we run a regression over the dichotomic variables indicating the presence of the pandemics and the consequent limitation in crowd attendance, using controls on the outcome of the game.\\

\section{Results}
 In Tables \ref{tab:mean1} and \ref{tab:mean2}, we can observe the result of applying a difference of means test. On the first case, over the \textit{NotHomeW} variable, we see that the one sided test rejects the null hypothesis of equality of means, with a $10\%$ of significance. On average, a $5\%$ more of away victories or draws can be seen ($36\%$ pre-pandemics against $41\%$ after). \\
 
 If we run a similar test on the difference of points in the match results, it rejects the null hypothesis, with a $5\%$ of significance in the case where we consider separately the dummy variables \textit{Covid} and \textit{Restricted} and with a $10\%$ in the case of \textit{Closed}. On average, home teams get $2$ points less.\\
 
\begin{table}[hbt!]
	\centering
	\caption{Difference of means test on visitor winnings}
\begin{tabular}{r r l r l r l}
	\hline
	& \multicolumn{2}{c}{Test 1} & \multicolumn{2}{c}{Test 2}
	& \multicolumn{2}{c}{Test 3} \\[1ex]
	
	Mean: & \multicolumn{2}{c}{{\em NotHomeW}} & \multicolumn{2}{c}{{\em NotHomeW}}
	& \multicolumn{2}{c}{{\em NotHomeW}} \\[1ex]
	
	\multirow{ 2}{*}{Condition:} & \multicolumn{2}{c}{{\em Covid}$=1$ vs.} & \multicolumn{2}{c}{{\em Restricted}$=1$ vs. }
	& \multicolumn{2}{c}{{\em Closed}$=1$ vs. } \\[1ex]
	& \multicolumn{2}{c}{{\em Covid}$=0$} & \multicolumn{2}{c}{ {\em Restricted}$=0$}
	& \multicolumn{2}{c}{ {\em Closed}$=0$} \\[1ex]
	\hline
	\hline
	\multirow{ 3}{*}{Mean (group 1)} & n & $565$ & n & $474$ & n & $370$  \\
	& mean & $0.4106$ & mean & $0.4135$ & mean & $0.4162$ \\
	& sd & $0.0207$ & sd & $0.0226$ & sd & $0.0256$ \\
	\hline
	\multirow{ 3}{*}{Mean (group 2)} & n & $462$ & n & $553$ & n & $657$ \\
	& mean & $0.3614$ & mean & $0.3671$ & mean & $0.3729$ \\
	& sd & $0.0223$ & sd & $0.0205$ & sd & $0.0188$ \\
	\hline
	\emph{Two tails p-value} &  & $0.1081$ &  & $0.1284$ &  & $0.1719$ \\
	\emph{One tail p-value} &  & $0.0541$ &  & $0.0642$ &  & $0.0860$ \\
	\hline
\end{tabular}
	\label{tab:mean1}
\end{table}	

\begin{table}[hbt!]
	\centering
	\caption{Difference of means test on local difference in points}
\begin{tabular}{r r l r l r l}
	\hline
	& \multicolumn{2}{c}{Test 1} & \multicolumn{2}{c}{Test 2}
	& \multicolumn{2}{c}{Test 3} \\[1ex]
	
	Mean: & \multicolumn{2}{c}{{\em DifLoc}} & \multicolumn{2}{c}{{\em DifLoc}}
	& \multicolumn{2}{c}{{\em DifLoc}} \\[1ex]
	
	\multirow{ 2}{*}{Condition:} & \multicolumn{2}{c}{{\em Covid}$=1$ vs.} & \multicolumn{2}{c}{{\em Restricted}$=1$ vs. }
	& \multicolumn{2}{c}{{\em Closed}$=1$ vs. } \\[1ex]
	& \multicolumn{2}{c}{{\em Covid}$=0$} & \multicolumn{2}{c}{ {\em Restricted}$=0$}
	& \multicolumn{2}{c}{ {\em Closed}$=0$} \\[1ex]
	\hline
	\hline
	\multirow{ 3}{*}{Mean (group 1)} & n & $565$ & n & $474$ & n & $370$  \\
	& mean & $4.626$ & mean & $4.443$ & mean & $4.367$ \\
	& sd & $0.786$ & sd & $0.891$ & sd & $1.056$ \\
	\hline
	\multirow{ 3}{*}{Mean (group 2)} & n & $462$ & n & $553$ & n & $657$ \\
	& mean & $6.712$ & mean & $6.526$ & mean & $6.238$ \\
	& sd & $0.974$ & sd & $0.852$ & sd & $0.758$ \\
	\hline
	\emph{Two tails p-value} &  & $0.0925$ &  & $0.0922$ &  & $0.1453$ \\
	\emph{One tail p-value} &  & $0.0463$ &  & $0.0461$ &  & $0.0726$ \\
	\hline
\end{tabular}
	\label{tab:mean2}
\end{table}

To control these differences with dummy variables that take into account the intrinsic characteristics of each game, we run a regression\footnote{Tables 4 to 7 can be found in Appendix 3}. Firstly, in Table \ref{tab:reg0} we include the variable \textit{Predic}, that predicts the chance of a win for the home team. Its high significance does not get reduced due to the presence of the Covid and crowd restriction variables, except in the case in which the points difference and the closure variables interact (with a p-value above the $10\%$ threshold).\\
 
Due to the presence of heterokedasticity, shown by White's test, we proceed to estimate the regressions with robust errors, controlling for the dummy variables corresponding to the country and the tournament. On Table \ref{tab:reg1}, we see that all the variables already discussed are still significant, other than the one distinguishing between partial and full closure.\\

The rest of the control variables have in general the expected signs and, except for  \textit{Travel} and \textit{StreakL}, are all significant. It is also interesting to find that the \textit{Gender} variable is significant, meaning that in women's tournaments the home advantage (measured as the winning rate) is not as relevant as in men's tournaments.\\ 

Table \ref{tab:reg2} presents the results for similar specifications of the controls, but with the points difference as the dependent variable. We obtain the same significance of the dichotomous variables, with patterns analogous to those in Table \ref{tab:reg1}.\\

A final  extension, suggested by the nature of the independent variable \textit{NotHomeW}, is to run a logistic regression. Its results can be found in Table \ref{tab:reg3}. Under this specification, the same general conclusions remain valid.\\

A final observation is that the coefficients obtained under the different specifications of the regression are stable under a test of differences in means.\\

\section{Conclusions}
In this paper we study the effect of crowd attendance on the home advantage in professional and semi-professional rugby. The unique opportunity created by the lockdowns due to Covid-19 was used to investigate this effect in a larger dataset than ever before. Despite the home advantage effect has been already established in several studies, it was difficult to isolate a dominant factor responsible for it. Our work intends to show how sport fans attending to the games may affect the final outcome, at least in rugby. \\

As a first result, we find that the home advantage (measured as percentage of home wins) decreases approximately $5\%$ when we consider full or partial restrictions. We also find that the point difference in favor of home teams decreases on average from approximately six to four points. This is an interesting result, since it puts away teams at just one try of difference of winning the game (a try is worth five points). So, a sports planner can use this information if she wants to induce a more exciting game in which the final outcome remains undecided until the end of the game.\\

Further analysis is needed to disentangle how the crowd attendance might affect the home advantage either through the exertion of social pressure on the referees  or by encouraging the home team (or even demotivating the away team).

\section*{References}
\begin{itemize}
\item Agnew, G. A., \& Carron, A. V. (1994). Crowd effects and the home advantage. International Journal of Sport Psychology, 25(1), 53–62.
\item Aguilar, Ó. G., Romero, J. J. F., \& García, M. S. (2014). Determination of the home advantage in handball Olympic Games and European Championships. Journal of Human Sport and Exercise, 9(4), 752-760.
\item Balmer, N. J., Nevill, A. M., \& Williams, A. M. (2001). Home advantage in the Winter Olympics (1908-1998). Journal of sports sciences, 19(2), 129-139.
\item Balmer, N. J., Nevill, A. M., \& Williams, A. M. (2003). Modelling home advantage in the Summer Olympic Games. Journal of sports sciences, 21(6), 469-478.
\item Belchior, C. A. (2020). Fans and Match Results: Evidence From a Natural Experiment in Brazil. Journal of Sports Economics, 21(7), 663-687.
\item Boudreaux, C. J., Sanders, S. D., \& Walia, B. (2017). A natural experiment to determine the crowd effect upon home court advantage. Journal of Sports Economics, 18(7), 737-749.
\item Boyko, R. H., Boyko, A. R., \& Boyko, M. G. (2007). Referee bias contributes to home advantage in English Premiership football. Journal of sports sciences, 25(11), 1185-1194.
\item Brown Jr, T. D., Van Raalte, J. L., Brewer, B. W., Winter, C. R., Cornelius, A. E., \& Andersen, M. B. (2002). World Cup Soccer Home Advantage. Journal of Sport Behavior, 25(2).
\item Bryson, A., Dolton, P., Reade, J. J., Schreyer, D., \& Singleton, C. (2021). Causal effects of an absent crowd on performances and refereeing decisions during Covid-19. Economics Letters, 198, 109664.
\item Carmichael, F., \& Thomas, D. (2005). Home-field effect and team performance: evidence from English premiership football. Journal of sports economics, 6(3), 264-281.
\item Cueva, C. (2020). Animal Spirits in the Beautiful Game. Testing social pressure in professional football during the COVID-19 lockdown. Working paper.
\item Downward, P., \& Jones, M. (2007). Effects of crowd size on referee decisions: Analysis of the FA Cup. Journal of sports sciences, 25(14), 1541-1545.
\item Endrich, M., \& Gesche, T. (2020). Home-bias in referee decisions: Evidence from “Ghost Matches” during the Covid19-Pandemic. Economics Letters, 197, 109621.
\item Ferreira Julio, U., Panissa, V. L. G., Miarka, B., Takito, M. Y., \& Franchini, E. (2013). Home advantage in judo: A study of the world ranking list. Journal of Sports Sciences, 31(2), 212-218.
\item García, M. S., Aguilar, Ó. G., Vázquez Lazo, C. J., Marques, P. S., \& Fernández Romero, J. J. (2013). Home advantage in Home Nations, Five Nations and Six Nations rugby tournaments (1883-2011). International Journal of Performance Analysis in Sport, 13(1), 51-63.
\item Jamieson, J. P. (2010). The home field advantage in athletics: A meta‐analysis. Journal of Applied Social Psychology, 40(7), 1819-1848.
\item Jones, M. V., Bray, S. R., \& Olivier, S. (2005). Game location and aggression in rugby league. Journal of Sports Sciences, 23(4), 387-393.
\item Jones, M. B. (2007). Home advantage in the NBA as a game-long process. Journal of Quantitative Analysis in Sports, 3(4).
\item Krumer, A. (2017). On winning probabilities, weight categories, and home advantage in professional judo. Journal of Sports Economics, 18(1), 77-96.
\item Legaz-Arrese, A., Moliner-Urdiales, D., \& Munguía-Izquierdo, D. (2013). Home advantage and sports performance: evidence, causes and psychological implications. Universitas Psychologica, 12(3), 933-943.
\item McCarrick, D., Bilalic, M., Neave, N., \& Wolfson, S. (2020). Home Advantage during the COVID-19 Pandemic in European football. Working paper.
\item McCutcheon, L. E. (1984). The home advantage in high school athletics. Journal of Sport Behavior, 7(4), 135.
\item Neave, N., \& Wolfson, S. (2003). Testosterone, territoriality, and the ‘home advantage’. Physiology \& Behavior, 78(2), 269-275.
\item Nevill, A. M., Holder, R. L., Bardsley, A., Calvert, H., \& Jones, S. (1997). Identifying home advantage in international tennis and golf tournaments. Journal of Sports Sciences, 15(4), 437-443.
\item Page, L., \& Page, K. (2010). Evidence of referees' national favoritism in rugby. Paper provided by National Centre for Econometric Research in its series NCER Working Paper Series with, (62).
\item Page, K., \& Page, L. (2010). Alone against the crowd: Individual differences in referees’ ability to cope under pressure. Journal of Economic Psychology, 31(2), 192-199.
\item Pojskić, H., Šeparović, V., \& Užičanin, E. (2011). Modelling home advantage in basketball at different levels of competition. Acta Kinesiologica, 5(1), 25-30.
\item Ponzo, M., \& Scoppa, V. (2018). Does the home advantage depend on crowd support? Evidence from same-stadium derbies. Journal of Sports Economics, 19(4), 562-582.
\item Schwartz, B., \& Barsky, S. F. (1977). The home advantage. Social forces, 55(3), 641-661.
\item Taylor, J. B., Mellalieu, S. D., James, N., \& Shearer, D. A. (2008). The influence of match location, quality of opposition, and match status on technical performance in professional association football. Journal of Sports Sciences, 26(9), 885-895.

\end{itemize}
\newpage

\section*{Appendix 1: Descriptive Statistics}

\begin{table}[hbt!]
	\centering
	\caption{Descriptive Statistics, observations 1 - 1027}
	\begin{tabular}{l c c c c c }
		Variable & Mean & Median & Std. Dev. & Min. & Max. \\[1ex]
		PointsL & 26.1 & 24.0 & 13.6 & 0.00 & 105\\
		PointsV & 20.6 & 19.0 & 11.1 & 0.00 & 74.0\\
		NotHomeW & 0.389 & 0.00 & 0.488 & 0.00 & 1.00\\
		DifLoc & 5.56 & 4.00 & 19.8 & $-$70.0 & 105\\
		Attendance & 1.18 & 2.00 & 0.932 & 0.00 & 2.00\\
		Predic & 0.307 & 0.00 & 0.461 & 0.00 & 1.00\\
		PosL & 7.17 & 7.00 & 4.22 & 1.00 & 18.0\\
		PosV & 7.22 & 7.00 & 4.18 & 1.00 & 22.0\\
		StreakL & 0.785 & 0.00 & 1.35 & 0.00 & 11.0\\
		StreakV & 0.893 & 0.00 & 1.35 & 0.00 & 10.0\\
		WinL & 2.46 & 2.00 & 2.52 & 0.00 & 16.0\\
		WinV & 2.54 & 2.00 & 2.54 & 0.00 & 16.0\\
		Covid & 0.550 & 1.00 & 0.498 & 0.00 & 1.00\\
		Gender & 0.0944 & 0.00 & 0.293 & 0.00 & 1.00\\
		Restricted & 0.462 & 0.00 & 0.499 & 0.00 & 1.00\\
		Closed & 0.360 & 0.00 & 0.480 & 0.00 & 1.00\\
		Travel & 0.1577 & 0.00  & 0.3647 & 0.00 & 1.00\\
	\end{tabular}
	\label{tab:stats}
\end{table}	
\newpage
\section*{Appendix 2: Histograms}
\begin{figure}[hbt!]
	\begin{subfigure}{.8\textwidth}
		\centering
	\caption{Covid=0}
	\includegraphics[width=\linewidth]{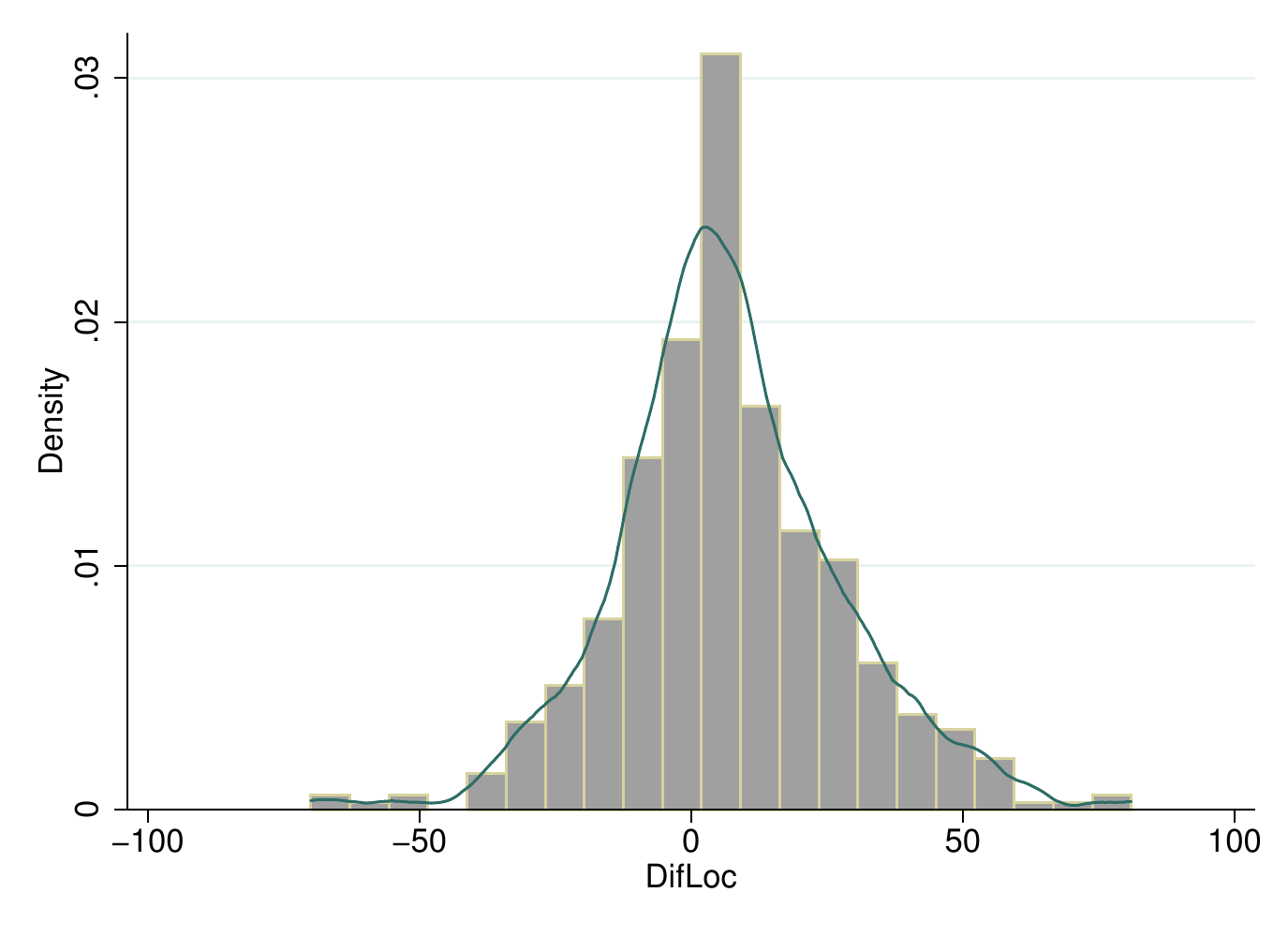}
	\label{fig:covid0}
    \end{subfigure}
\newline
\begin{subfigure}{.8\textwidth}
	\centering
	\caption{Covid=1}
	\includegraphics[width=\linewidth]{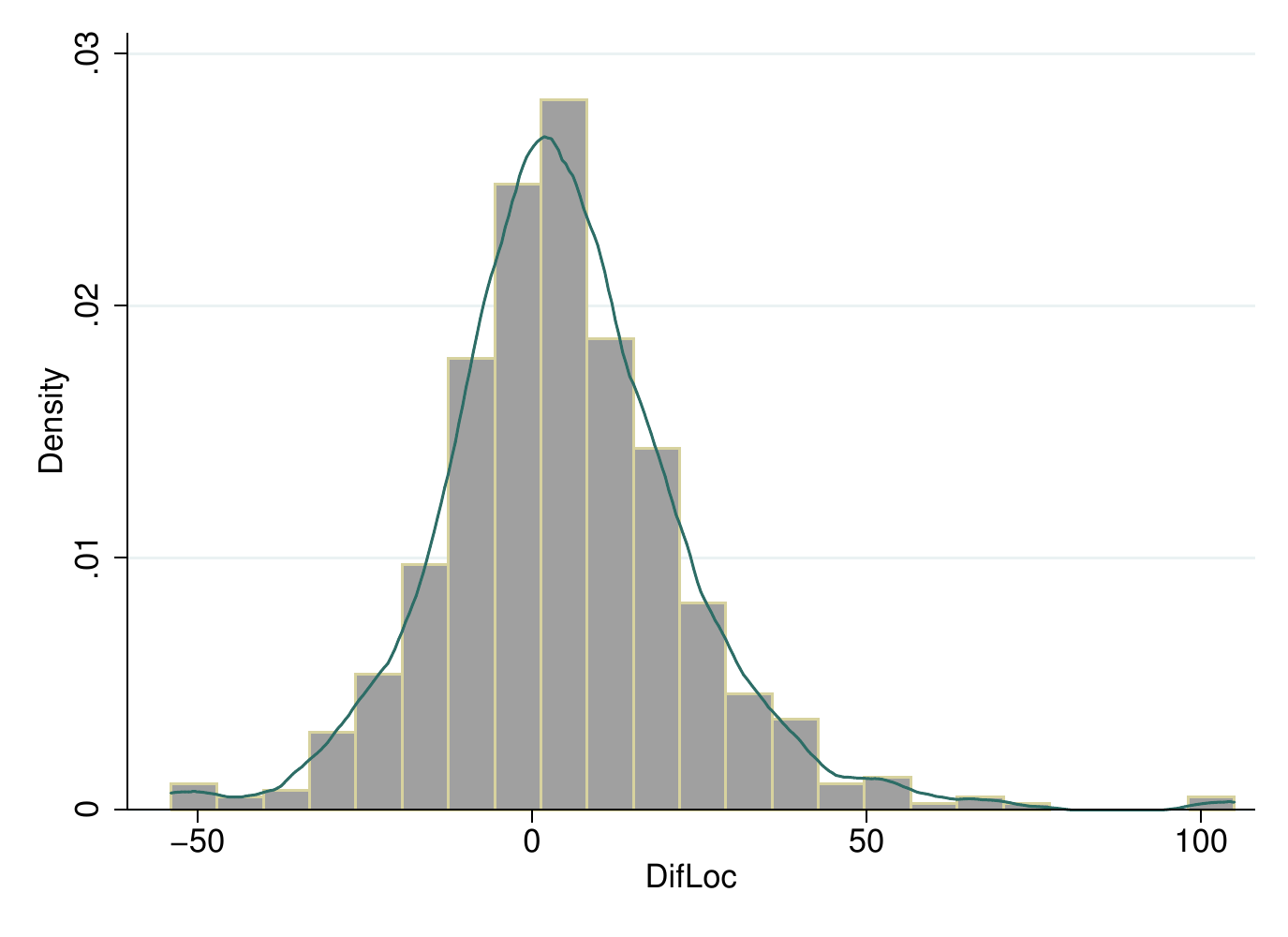}
	\label{fig:covid1}
\end{subfigure}
\end{figure}
\newpage

\newpage

\section*{Appendix 3: Tables}
\begin{table}[hbt!]
	\centering
	\caption{Regressions}
	\begin{tabular}{lcccccc} \hline
		& (1) & (2) & (3) & (4) & (5) & (6) \\
		VARIABLES & NotHomeW & NotHomeW & NotHomeW & DifLoc & DifLoc & DifLoc \\ \hline
		&  &  &  &  &  &  \\
		Predic & 0.532*** & 0.532*** & 0.529*** & -11.51*** & -11.52*** & -11.43*** \\
		& (0.0325) & (0.0325) & (0.0325) & (1.388) & (1.388) & (1.388) \\
		Closed & 0.0572* &  &  & -2.018 &  &  \\
		& (0.0304) &  &  & (1.300) &  &  \\
		Restricted &  & 0.0624** &  &  & -2.358* &  \\
		&  & (0.0300) &  &  & (1.284) &  \\
		Covid &  &  & 0.0554* &  &  & -2.255* \\
		&  &  & (0.0315) &  &  & (1.344) \\
		Constant & 0.204*** & 0.194*** & 0.192*** & 9.866*** & 10.31*** & 10.48*** \\
		& (0.0221) & (0.0244) & (0.0272) & (0.944) & (1.044) & (1.164) \\
		&  &  &  &  &  &  \\
		Observations & 796 & 796 & 796 & 796 & 796 & 796 \\
		R-squared & 0.254 & 0.255 & 0.254 & 0.082 & 0.083 & 0.082 \\ \hline
		\multicolumn{7}{c}{ Standard errors in parentheses} \\
		\multicolumn{7}{c}{ *** p$<$0.01, ** p$<$0.05, * p$<$0.1} \\
	\end{tabular}
	\label{tab:reg0}
\end{table}	

\begin{landscape}
	\scriptsize
	\begin{table}[!htbp] \centering 
		\caption{NotHomeW Results}

\begin{tabular}{lcccccc} \hline
 & (1) & (2) & (3) & (4) & (5) & (6) \\
VARIABLES & NotHomeWin & NotHomeWin & NotHomeWin & NotHomeWin & NotHomeWin & NotHomeWin \\ \hline
 &  &  &  &  &  &  \\
Closed & 0.0459 &  &  & 0.0446 &  &  \\
 & (0.0355) &  &  & (0.0339) &  &  \\
 Restricted &  & 0.0539* &  &  & 0.0496 &  \\
 &  & (0.0320) &  &  & (0.0308) &  \\
 Covid &  &  & 0.0622** &  &  & 0.0556* \\
 &  &  & (0.0308) &  &  & (0.0297) \\
WinL & -0.0631*** & -0.0632*** & -0.0634*** &  &  &  \\
 & (0.00846) & (0.00845) & (0.00843) &  &  &  \\
WinV & 0.0647*** & 0.0650*** & 0.0654*** &  &  &  \\
 & (0.00803) & (0.00798) & (0.00795) &  &  &  \\
Gender & 0.303* & 0.306* & 0.312* &  &  &  \\
 & (0.168) & (0.168) & (0.167) &  &  &  \\
Travel & -0.0266 & -0.0177 & -0.00744 &  &  &  \\
 & (0.0809) & (0.0814) & (0.0820) &  &  &  \\
PosL &  &  &  & 0.0215*** & 0.0215*** & 0.0215*** \\
 &  &  &  & (0.00378) & (0.00378) & (0.00377) \\
PosV &  &  &  & -0.0198*** & -0.0197*** & -0.0196*** \\
 &  &  &  & (0.00366) & (0.00366) & (0.00365) \\
StreakL &  &  &  & -0.0136 & -0.0139 & -0.0138 \\
 &  &  &  & (0.0117) & (0.0117) & (0.0117) \\
StreakV &  &  &  & 0.0370*** & 0.0371*** & 0.0376*** \\
 &  &  &  & (0.0115) & (0.0115) & (0.0115) \\
 \hline
Country & YES & YES & YES & YES & YES & YES \\
Tournament & NO & NO & NO & YES & YES & YES  \\
Constant & 0.286*** & 0.278*** & 0.267*** & 0.260*** & 0.254*** & 0.244*** \\
 & (0.0407) & (0.0412) & (0.0418) & (0.0561) & (0.0564) & (0.0573) \\
 &  &  &  &  &  &  \\
Observations & 1,027 & 1,027 & 1,027 & 1,027 & 1,027 & 1,027 \\
 R-squared & 0.117 & 0.118 & 0.119 & 0.129 & 0.130 & 0.131 \\ \hline
\multicolumn{7}{c}{ Robust standard errors in parentheses} \\
\multicolumn{7}{c}{ *** p$<$0.01, ** p$<$0.05, * p$<$0.1} \\
\end{tabular}

		\label{tab:reg1}
	\end{table}  
	
\end{landscape}

\begin{landscape}
	\scriptsize
	\begin{table}[!htbp] \centering 
		\caption{DifLoc Results}

\begin{tabular}{lcccccc} \hline
 & (1) & (2) & (3) & (4) & (5) & (6) \\
VARIABLES & DifLoc & DifLoc & DifLoc & DifLoc & DifLoc & DifLoc \\ \hline
 &  &  &  &  &  &  \\
Closed & -2.492* &  &  & -2.065 &  &  \\
 & (1.481) &  &  & (1.341) &  &  \\
 Restricted &  & -2.574** &  &  & -2.236* &  \\
 &  & (1.275) &  &  & (1.204) &  \\
 Covid &  &  & -2.694** &  &  & -2.341** \\
 &  &  & (1.231) &  &  & (1.172) \\
WinL & 3.664*** & 3.660*** & 3.661*** &  &  &  \\
 & (0.358) & (0.357) & (0.356) &  &  &  \\
WinV & -3.350*** & -3.374*** & -3.395*** &  &  &  \\
 & (0.337) & (0.331) & (0.329) &  &  &  \\
Gender & -12.05 & -12.27 & -12.58 &  &  &  \\
 & (7.704) & (7.690) & (7.693) &  &  &  \\
Travel & 4.482 & 4.098 & 3.722 &  &  &  \\
 & (2.895) & (2.889) & (2.903) &  &  &  \\
PosL &  &  &  & -1.087*** & -1.089*** & -1.090*** \\
 &  &  &  & (0.143) & (0.143) & (0.143) \\
PosV &  &  &  & 1.093*** & 1.091*** & 1.087*** \\
 &  &  &  & (0.134) & (0.133) & (0.133) \\
StreakL &  &  &  & 1.442*** & 1.453*** & 1.446*** \\
 &  &  &  & (0.462) & (0.462) & (0.461) \\
StreakV &  &  &  & -1.887*** & -1.891*** & -1.915*** \\
 &  &  &  & (0.408) & (0.407) & (0.405) \\
 \hline
  Country & YES & YES & YES & YES & YES & YES \\
 Tournament & YES & YES & YES & YES & YES & YES  \\
Constant & 8.721*** & 9.027*** & 9.441*** & 9.778*** & 10.04*** & 10.41*** \\
 & (1.410) & (1.431) & (1.469) & (1.960) & (1.963) & (1.994) \\
 &  &  &  &  &  &  \\
Observations & 1,027 & 1,027 & 1,027 & 1,027 & 1,027 & 1,027 \\
 R-squared & 0.179 & 0.180 & 0.180 & 0.198 & 0.199 & 0.199 \\ \hline
\multicolumn{7}{c}{ Robust standard errors in parentheses} \\
\multicolumn{7}{c}{ *** p$<$0.01, ** p$<$0.05, * p$<$0.1} \\
\end{tabular}

		\label{tab:reg2} 	
	\end{table}  
	
\end{landscape}

\begin{landscape}
	\scriptsize
	\begin{table}[!htbp] \centering 
		\caption{Logit Results}

\begin{tabular}{lccccccccc} \hline
 & (1) & (2) & (3) & (4) & (5) & (6) & (7) & (8) & (9) \\
VARIABLES & NotHomeWin & NotHomeWin & NotHomeWin & NotHomeWin & NotHomeWin & NotHomeWin & NotHomeWin & NotHomeWin & NotHomeWin \\ \hline
 &  &  &  &  &  &  &  &  &  \\
Closed & 0.329* &  &  & 0.204 &  &  & 0.233 &  &  \\
 & (0.193) &  &  & (0.154) &  &  & (0.208) &  &  \\
Restricted &  & 0.383** &  &  & 0.225 &  &  & 0.355* &  \\
&  & (0.178) &  &  & (0.141) &  &  & (0.200) &  \\
Covid &  &  & 0.483** &  &  & 0.285** &  &  & 0.493** \\
&  &  & (0.194) &  &  & (0.139) &  &  & (0.224) \\
PosL &  &  &  & 0.106*** & 0.106*** & 0.107*** &  &  &  \\
 &  &  &  & (0.0179) & (0.0179) & (0.0179) &  &  &  \\
PosV &  &  &  & -0.0951*** & -0.0953*** & -0.0947*** &  &  &  \\
 &  &  &  & (0.0178) & (0.0178) & (0.0178) &  &  &  \\
StreakL &  &  &  & -0.109** & -0.110** & -0.111** &  &  &  \\
 &  &  &  & (0.0553) & (0.0554) & (0.0555) &  &  &  \\
StreakV &  &  &  & 0.170*** & 0.170*** & 0.173*** &  &  &  \\
 &  &  &  & (0.0522) & (0.0521) & (0.0521) &  &  &  \\
Predic & 2.448*** & 2.459*** & 2.457*** &  &  &  & 2.510*** & 2.523*** & 2.524*** \\
 & (0.195) & (0.196) & (0.195) &  &  &  & (0.191) & (0.192) & (0.192) \\
Gender &  &  &  &  &  &  & 2.052 & 2.028 & 2.036 \\
 &  &  &  &  &  &  & (1.429) & (1.434) & (1.438) \\
Travel &  &  &  &  &  &  & -0.245 & -0.172 & -0.0662 \\
 &  &  &  &  &  &  & (0.406) & (0.410) & (0.417) \\
 \hline
Country & YES & YES & YES & NO & NO & NO & YES & YES & YES \\
Tournament & NO & NO & NO & YES & YES & YES & YES & YES & YES \\
Constant & -1.677*** & -1.758*** & -1.886*** & -1.110*** & -1.132*** & -1.205*** & -1.611*** & -1.686*** & -1.805*** \\
 & (0.176) & (0.191) & (0.220) & (0.273) & (0.275) & (0.281) & (0.221) & (0.228) & (0.247) \\
 &  &  &  &  &  &  &  &  &  \\
 Observations & 796 & 796 & 796 & 1,027 & 1,027 & 1,027 & 796 & 796 & 796 \\ \hline
\multicolumn{10}{c}{ Robust standard errors in parentheses} \\
\multicolumn{10}{c}{ *** p$<$0.01, ** p$<$0.05, * p$<$0.1} \\
\end{tabular}

		\label{tab:reg3}
	\end{table}  
	
\end{landscape}
\end{document}